\documentclass[amsmath,amssym,amsfonts,pre,twocolumn,showpacs]{revtex4}
\usepackage[dvips]{epsfig}

\begin{document}

\newcommand{\Xv}{{\mathbf x}}
\newcommand{\qv}{{\mathbf q}}
\newcommand{\av}{{\mathbf a}}
\newcommand{\cH}{{\cal H}}
\newcommand{\lB}{\ell_B}
\newcommand{\apa}{a_\parallel}
\newcommand{\ape}{a_\perp}
\newcommand{\Sv}{\hat{{\bf S}}}

\newcommand{\tv}{\hat{{\bf t}}}
\newcommand{\bv}{{\bf b}}
\newcommand{\xv}{{\mathbf x}}
\newcommand{\uv}{{\mathbf u}}
\newcommand{\vv}{{\mathbf v}}
\newcommand{\pv}{{\mathbf p}}
\newcommand{\rv}{{\mathbf r}}
\newcommand{\Rv}{{\mathbf R}}
\newcommand{\ev}{ {{\bf e}} }
\newcommand{\Tv}{ {{\bf T} }}
\newcommand{\Nv}{{{\bf N}}}
\newcommand{\Bv}{{{\bf B}}}

\title{Frustration and Packing in Curved-Filament Assemblies:  From Isometric to Isomorphic Bundles}

\author{Gregory M. Grason}
\affiliation{Department of Polymer Science and Engineering, University of Massachusetts, Amherst, MA 01003, USA}

\begin{abstract}
Densely-packed bundles of biological filaments (filamentous proteins) are common and critical structural elements in range of biological materials.  While most bundles form from intrinsically straight filaments, there are notable examples of protein filaments possessing a natural, or intrinsic, curvature, such as the helical bacterial flagellum.  We study the non-linear interplay between thermodynamic preference for dense and regular inter-filament packing and the mechanical preference for uniform filament shape in bundles of helically-curved filaments.  Geometric constraints in bundles make perfect inter-filament (constant spacing, or {\it isometric}) packing incompatible with perfect intra-filament (constant shape, or {\it isomorphic}) packing.  As a consequence, we predict that bundle packing exhibits a strong sensitivity to bundle size, evolving from the isometric packing at small radii to an isomorphic packing at large radii.  The nature of the transition between these extremal states depends on thermodynamic costs of packing distortion, with packing in elastically-constrained bundles evolving smoothly with size, while packing in osmotically-compressed bundles may exhibit a singular transition from the isometric packing at a finite bundle radius.  We consider the equilibrium assembly of bundles in a saturated solution of filaments and show that mechanical cost of isomorphic packing leads to self-limited equilibrium bundle diameters, whose size and range of thermodynamic stability depend both on condensation mechanism, as well as the helical geometry of filaments.
\end{abstract}
\date{\today}

\maketitle

\section{Introduction}   

Molecular filaments and their assemblies underlie the function of a diverse range of materials, from intra- and inter-cellular matter to high-strength synthetic materials.  Condensed, parallel arrays, or bundles, of cytoskelatal filaments like f-actin and microtubules provide both mechanical reinforcement to cells and play crucial roles in vital cellular processes like adhesion, division and locomotion~\cite{theriot, popp}.  Rope-like assemblies of extra-cellular protein filaments --- for example, collagen, cellulose or fibrin --- are also primary structural elements {\it in-between} cells, in plant and animal tissue~\cite{fratzl, weisel}.  As filaments are densely-packed and hence strongly interacting in bundles, the structure of intermolecular organization underlies the collective dynamic and mechanical properties of these assemblies, critical for their robust function in living systems.  In the simplest models of this material prototype, the ground state of bundles of straight filaments is universally a hexagonally-packed array of parallel filaments, which maximizes both the number of inter-filament contacts as well as the density within the bundle.  As a consequence, filament bundles forming under the influence of cohesive inter-filament forces~\cite{wong} as well as osmotic pressure (depletion forces)~\cite{needleman} are observed to form hexagonal cross-sectional order.  Notably both the (straight) filament shape and the inter-filament spacing are uniform throughout a hexagonal bundle, which implies that both length and lateral size of self-assembled bundles is thermodynamically unlimited~\cite{grason_prl_07, grason_pre_09}.

\begin{figure}
\epsfig{file=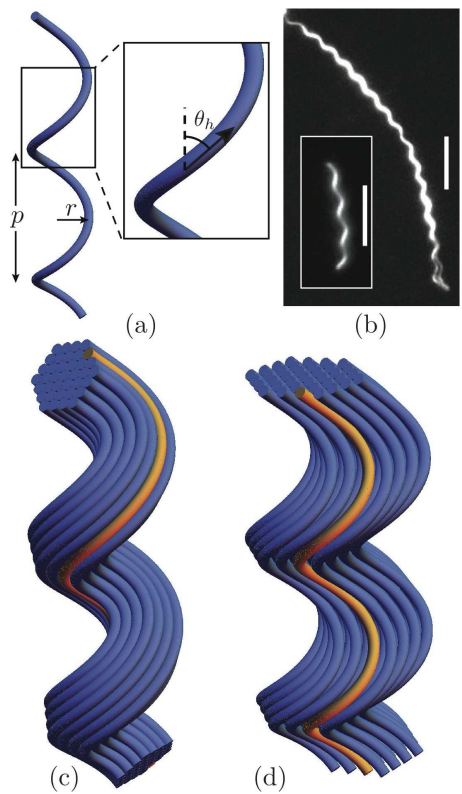, width=2.35in}
\caption{Preferred (stress free) geometry of helical filaments shown in (a), where $\tan \theta_h = 2 \pi r/p$.  In (b), florescence-microscopy images of a single bacterial flagella (inset) and a bundle of ($\sim 10$) flagella formed in a neutral polymer suspension, scale bars are 5 $\rm \mu m$ (courtesy of Z. Dogic, Brandeis University).  The two extreme cases of bundle packing of helical filaments:  (c) the {\it isometric} (constant spacing) packing; and (d) the {\it isomorphic} (constant shape) packing.  The orange filament on the surface of the bundles highlights the subtle changes of filament shape required by the isometric packing in (c), while the lateral spacing in (d) is such that filament diameters do not overlap in the bundle interior. }
\label{figure1}
\end{figure}

In this article, we consider the more complex structure and thermodynamics of bundles formed by {\it intrinsically-curved} filaments, a distinction which we show frustrates homogeneous inter- and intra-filament geometry of straight, hexagonal bundles.  For many common biofilaments --- f-actin or microtubules --- the filament backbone is straight in the absence of external stress or thermal fluctuations.  There are, however, important exceptions, where the backbone filament geometry has a preferred, or natural, curvature, deriving from the shape and interactions between globular subunits~\cite{asakura_oosawa_book}.  Most widely known among this class of intrinsically-curved filaments is the bacterial {\it flagella}~\cite{namba}, a corkscrew shaped protein filament---helical radius $r =0.2 {\rm \mu m}$; and pitch $p=2.2 {\rm \mu m}$; diameter $\approx 20$ nm, shown in Fig.~\ref{figure1} (b)---that propels common microorganisms like {\it E. coli} and {\it salmonella}.   Notably, bacteria swim, not by the actuation of single filaments, but by the screw-like rotation of coherent bundles of multiple flagella~\cite{macnab}, held together through a combination of contact and hydrodynamic interactions.  More recently, new classes of curved-filament assemblies have been discovered as components of bacterial ``cytoskeleton"~\cite{popp}.  One such filament MReb, considered to be a prokaryotic analogue to f-actin, is observed to associate with inner-cell wall of bacteria in bundles that wind helically around the cell body, potentially shaping the growth of rod-like bacteria~\cite{andrews, margolin}.  A second filament, crescentin, also forms bundles {\it in vivo}~\cite{jensen} whose helical structure is believed to drive the curved and helical shapes of bacteria like {\it Caulabacter crescentus}~\cite{ausmees}.

Unlike the case of straight-filament assembly, bundles of curved-filaments are subject to an unavoidable geometric conflict between inter-filament packing and the natural curvature of filaments.  Simply put, it is not possible to evenly space nearest-neighbor filaments throughout the ordered cross-section of the bundle while maintaining uniform shape of all filaments.  This conflict is illustrated by the bundles of helical filaments in Fig.~\ref{figure1} (c) and (d), which we denote, respectively, as {\it isometric} (constant spacing) and {\it isomorphic} (constant shape) packings.

The perfect inter-filament packing --- constant center-to-center filament distance --- in the isometric bundle constrains the relative geometry between filaments so that the inter-filament separation, ${\bf \Delta}$, between any filament pair is {\it relatively parallel} to the filament tangents~\cite{starostin}.  This means that in isometric bundles all filament backbones are normal to the common plane of lattice packing, and ${\bf \Delta}$ does not {\it twist} around the backbones along the length of the bundle [Fig.~\ref{figure1} (c)].  Importantly, this packing constrains the relationship between the curvature $\kappa$ and torsion $\tau$ of a ``reference" filament and the shape of a filament at separation ${\bf \Delta}$, whose respective curvature and torsion follow~\cite{grason_pre_09}
\begin{equation}
\kappa ({\bf \Delta})= \frac{ \kappa}  { 1- \kappa  (\Nv \cdot  {\bf \Delta}) } ; \
\tau  ({\bf \Delta}) = \frac{ \tau}  { 1- \kappa (\Nv \cdot  {\bf \Delta})} ,
\label{eq: kaptau}
\end{equation}
where $\Nv$ is the normal to reference filament.  Eq. (\ref{eq: kaptau}) is the filamentary analogue of the ``curvature focussing" properties of evenly-spaced, layered materials (i.e. smectics)~\cite{sethna, fournier, didonna, kleman}, embodying the fundamental frustration of the lateral assembly of filaments in densely-packed bundles.  On one hand, driving forces for dense-filament packing or cohesive filament generically penalize deviations from perfect lateral packing in a bundle.  However, when these filaments are intrinsically curved, meaning that $\kappa\neq 0$ in their stress free configurations, perfect lateral packing requires the filament geometry to vary along normal directions  ($\Nv \cdot {\bf \Delta} \neq 0$) in the bundle cross section, incurring a mechanical cost for change of filament shape.   Thus, in bundles of intrinsically-curved filaments, isometric packing is not compatible with constant filament shape, a property achieved in the isomorphic packing of Fig.~\ref{figure1} (d).  In the isomorphic state, as we show below, the non-local nature of filament contact requires inter-filament spacing to vary {\it along} the bundle, straining either interactions that prefer constant spacing or thermodynamic forces favoring maximal density.  The mutual frustration between isometric and isomorphic packing implies that the state of inter-filament packing in curved-filament bundle is determined not only by the lateral forces that drive filaments in to tightly-packed assemblies, but also by the mechanical costs of filament distortion. Here, we show that interplay between geometry and mechanics in curved-filament bundles has a profound influence on the structure and dimensions of assembly, giving rise to thermodynamic behavior not exhibited by assemblies of unfrustrated, straight filaments.

In this paper, we study the frustration between constant spacing and constant shape in bundles of helically-shaped bundles, whose backbones have preferred values of curvature and torsion, $\kappa_0$ and $\tau_0$.  Helical filaments are prototypes of the frustration in curved-filament assemblies, whose degree of frustration can be tuned continuously from zero for straight filaments ($\kappa_0/\tau_0 \ll 1$) and the maximal frustration of ring filaments ($\kappa_0/\tau_0 \gg 1)$.  In ref.~\cite{grason_pre_09}, it was shown that the mechanical costs of filament shape change in isometric, helical filament bundles grow rapidly with lateral size of bundles, ultimately limiting the lateral radius of bundles that are unable to relax the constant-spacing constraints.  In the present study, we develop a continuum theory to analyze the non-linear interplay between the distinct costs of inter-filament and intra-filament distortions in helical-filament bundles which relax both shape as well as the lateral packing.  We study two models for the thermodynamic cost of inter-filament distortion:  1) an elastic model for disruption of inter-filament cohesion; and 2) the osmotic excluded-volume cost of bundles formed in concentrated solutions of osmolytes excluded from the filament packing.  Analyzing a class of low-energy bundles that span continuously from the limiting isometric to isomorphic bundles, we show that bundle packing evolves from perfect inter-filament spacing in the limit of vanishing bundle diameter to constant filament shape in the limit of large diameter.  While inter-filament packing in cohesive bundles is distorted for any finite radius, we find in the limit of weak resistance to filament torsion, osmotically-condensed bundles remain isometric over finite range of small bundle size.  Finally, the consider the equilibrium size of bundles in the canonical ensemble, showing that the mechanical cost of shape change in small bundles stabilizes finite size bundles when the surface energy of bundles is below a threshold value, which is shown to increase with $\kappa_0/\tau_0$ for both bundling mechanisms.

The remainder of this article is organized as follows.  In Sec. II.A we introduce the geometry of helical filament bundles, as well as the mechanical cost for filament distortion and the thermodynamic costs for distortion of the packing in Sec. II.B.  In Sec. III.A we analyze the variation of bundle packing with lateral size of bundles in both cohesive and osmotically-driven bundles, and in Sec. III.B we consider the equilibrium radius of bundles in solutions of freely-associating filaments above saturation.  Finally we conclude in Sec. IV with a discussion of the implications on our study for bundles of flagellar filaments.

\section{Structure and mechanics of inter-filament packing in bundles}   

We consider two models of helical-filament bundles, which differ in terms of the driving forces for assembly.  In the first model, assembly is driven by short-range, adhesive interactions between neighbor filaments in the bundle, while the second model considers densely packed bundles to form under the influence of osmotic pressure, that works against the total volume of the bundle interior.  In both models, an increase of inter-filament packing forces (adhesion or pressure) drives the bundles away from the isomorphic state preferred by unbundled filaments towards the isometric packing.  To describe this transition, we consider a class of bundle configurations that may be continuously transformed from the isomorphic into the isometric state and {\it vice versa}.

\subsection{Geometry of filament packing}

We consider bundle configurations described by a packing the cross-section around a central filament, whose center is described by, $\Rv_0(s)$.  The backbone of the central filament is described by a tangent, normal or bi-normal, the Frenet frame $\{\Tv, \Nv, \Bv\}$, curvature $\kappa$ and torsion $\tau$, deriving from the relations,  
\begin{equation}
\partial_s \Rv_0 = \Tv; \ \partial_s \Tv = \kappa \Nv; \  \partial_s \Nv = -\tau \Bv.
\end{equation}
The curvature and torsion of this filament define the {\it helical angle} $\theta_h$, 
\begin{equation}
\tan \theta_h =  \kappa / \tau, 
\end{equation}
shown in Fig. ~\ref{figure1} (a).  In the bundle cross-section, filaments are labeled by coordinates $(x_u,x_v)$.  The vector $\xv= x_u \uv + x_v \vv$ maps the position of the central filament to the filament at $(x_u,x_v)$ in the plane spanned by the basis vectors, $\uv$ and $\vv$, 
\begin{equation}
\Rv(x_u,x_v,s) = \Rv_0(s) + \xv .
\label{Rv}
\end{equation}
We refer to the plane normal to $\pv = \uv \times \vv$ as the {\it packing plane}, and the orthonormal frame $\{ \uv, \vv, \pv \}$ as the {\it packing frame}.  We assume a regular, hexagonal lattice packing, i.e. $x_u = a(n+m/2)$ and $x_v = a m \sqrt{3}/2$ where $m$ and $n$ are integers labeling occupied lattice positions and $a$ lattice spacing.  

Due to the constraints of packing, the geometry of filaments varies through the cross-section of a bundle.  Given our construction, we describe the variation in terms of the functions $\Tv(\xv)$, $\kappa(\xv)$ and $\tau(\xv)$ characterizing respective values of backbone orientation, curvature and torsion for filaments at $\xv$.  We focus on the limit of narrow bundle cross-sections, such that $\kappa |\xv| \ll1 $ and $\tau |\xv| \ll1 $.  The shape of the filament at $\xv$ derives from both the shape of $\Rv_0(s)$ as well as the tilt and rotation of the frame $\{ \uv, \vv, \pv \}$ with respect to the geometry of the central curve at $\xv=0$.  The rotation of filament positions in the packing plane along the arc length of the bundle is described by the transformation,
\begin{eqnarray}
\nonumber
\uv &=& \cos (\Omega s) \uv'+  \sin (\Omega s) \vv' \\
\vv &=&- \sin (\Omega s) \uv'+  \cos (\Omega s)\vv' , 
\label{uv}
\end{eqnarray}  
where
\begin{equation}
\uv' = \Nv ; \ \vv' = \sin \theta \Tv + \cos \theta \Bv; \ \pv = \cos \theta \Tv - \sin \theta \Bv .
\label{upvp}
\end{equation}
Here,  $\theta$ is the tilt angle of the packing plane, the {\it packing angle}, with respect to the central tangent, i.e. $\cos \theta = \Tv \cdot \pv$.  Thus, $\Omega$ and $\theta$ constitute two independent parameters specifying the geometry of the packing.  We consider bundles that are homogeneous along their length so that both $\Omega$ and $\theta$ are constant.

The relationship between geometry (i.e. orientation, curvature and torsion) of the central filament to the geometry of a filament at $\xv$ derives from rotation of the  $\{ \uv, \vv, \pv \}$ frame along the bundle, which follows straightforwardly from eqs. (\ref{uv}) and (\ref{upvp})
\begin{eqnarray}
\nonumber
\partial_s \uv &=& ( \Omega + \Omega_0) \vv - \Omega_p \cos (\Omega s)   \pv \\
\nonumber
\partial_s \vv &=&- ( \Omega +\Omega_0) \uv +  \Omega_p \sin (\Omega s)  \pv \\
\partial_s \pv &=& \Omega_p \uv'  =  \Omega_p \big[ \cos (\Omega s) \uv + \sin (\Omega s) \vv \big] ,
\label{rot}
\end{eqnarray}
where
\begin{eqnarray}
\nonumber
\Omega_0 &=& - \vv' \cdot \partial_s \uv' = -\kappa \sin \theta + \tau \cos \theta \\
\Omega_p &=& - \pv \cdot \partial_s \uv' = \kappa \cos \theta + \tau \sin \theta .
\label{omegas}
\end{eqnarray}
From eq. (\ref{rot}) we note that the net rotation of filament positions around the center of the cross section, $ \vv \cdot \partial_s \uv=(\Omega+\Omega_0)$ is the simply the sum rotation of $\uv'$ and $\vv'$ around $\pv$, $\Omega_0$, and the rotation of $\uv$ and $\vv$ relative the to that coordinate system, $\Omega$.  In our analysis, we focus on the set of low-energy states that span the two-limiting packings, isometric and isomorphic.  We show in the appendix that vanishing torsion of filament positions in the packing plane (i.e. $\Omega+\Omega_0 =0$) is a necessary condition for both constant filament shape and constant filament spacing, and we therefore, expect this condition to also describe the low-energy states intermediate to these extreme cases.  That is, we restrict our analysis to
\begin{equation}
\Omega= - \Omega_0 ,
\end{equation}
so that there is {\it no rotation} of the filament position in the packing plane along the bundle.  

The variation of filament geometry throughout the packing plane derives from a calculation of the Frenet geometry based on the construction of eqs. (\ref{uv})-(\ref{omegas}) (see Appendix).  For backbone orientation (tangents) we find, first order (in $\kappa |\xv|$) correction $\Tv(\xv) - \Tv(0) \simeq \delta \Tv_\perp $
\begin{eqnarray}
\nonumber
\delta \Tv_\perp &=& \partial_s \xv - \Tv (\partial_s \xv \cdot \Tv)\\ &=&  \sin \theta \Omega_p (\Nv \cdot \xv) \Bv  ,
\label{dTperp}
\end{eqnarray}
which is manifestly perpendicular to $\Tv$.  The variation of backbone curvature and torsion are described by
\begin{equation}
\kappa(\xv) \simeq \kappa + \delta \kappa ~(\kappa \Nv \cdot \xv) ; \ \tau(\xv) \simeq \tau+ \delta \tau  ~  (\kappa \Nv \cdot \xv) 
\label{kaptau}
\end{equation}
where
\begin{eqnarray}
\nonumber
\delta \kappa 
&= &\Omega_p \big( 2 \cos \theta  -  \Omega_p / \kappa \big) 
\\ & =&  \kappa (1- \sin^2 \theta/ \sin^2 \theta_p)
\label{dkappa}
\end{eqnarray}
and 
\begin{eqnarray}
\nonumber
\delta \tau &=&  \Omega_p  \big(\cos \theta~  \tau / \kappa + \sin \theta  - \sin \theta  \Omega_0^2/\kappa^2\big)  \\ \nonumber & = &  \kappa \cos^2 \theta ( \tan \theta \cot \theta_p - 1) \big[ \tan \theta - \cot \theta_p \\ && - \cos \theta \sin \theta ( \cot \theta_p + \tan \theta )^2\big] ,
\label{dtau}
\end{eqnarray}
where the common root of expressions occurs for packing angle,
\begin{equation}
\theta_p \equiv - \arctan(\kappa/\tau) = - \theta_h.
\end{equation}
Notably, shape change of filaments (to lowest order) is distributed anisotropically in the packing plane:  curvature and torsion are maximally distorted for filaments along $\xv \parallel \Nv$, while no shape change is required for filaments at $\xv \perp \Nv$.

\begin{figure*}
\epsfig{file=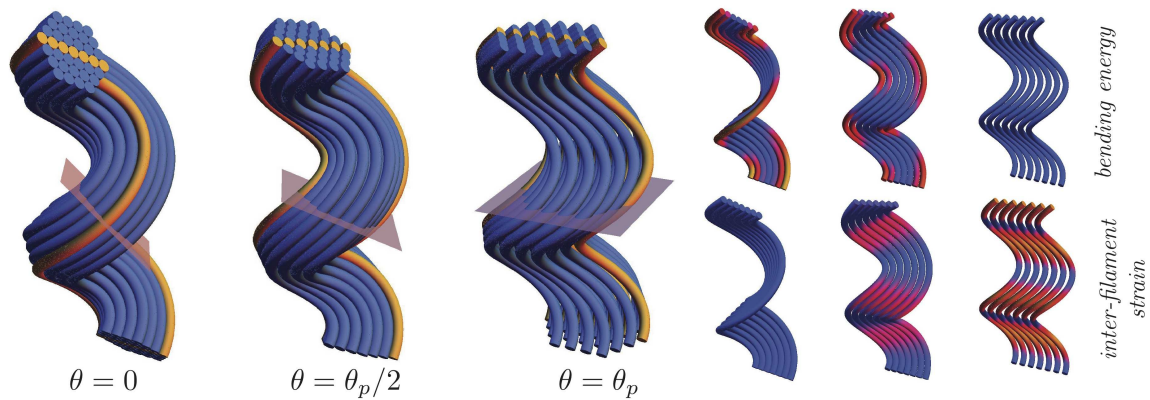, width=6in}
\caption{On the left, three examples of the helical filament packings considered in our model, spanning from isometric ($\theta =0$) to isomorphic ($\theta= \theta_p$).  The orientation of the ``packing plane" of hexagonal order is shown in each.  In each bundle, orange filaments highlight a single row of filaments, whose bending and inter-filament packing geometry is depicted on the right for the same three packing angles.  On the top right, the color variation highlights local deviations from preferred curvature, $|\kappa(\xv) - \kappa|^2$, ranging from blue for undistorted to yellow for highly-distorted curvature.  In the bottom right, the color variation highlights strain of interfilament distance along the row, deriving from the local tilt in the packing plane along the neighbor direction according to eq. (\ref{Delperp}), where blue indicates no strain and yellow indicates high strain.  }
\label{packing}
\end{figure*}

This construction encompasses two extremal limits of the packing geometry:

{\it Isomorphic Packing}: $ \Omega_p = 0$ - Inspection of eqs. (\ref{dTperp}) - (\ref{dtau}) immediately reveals these conditions achieve uniform orientation and shape the packing plane, $\delta \Tv = \delta \kappa = \delta \tau = 0$.  Solution of eq. (\ref{omegas}) gives the inclination and rotation rate, $\Omega$, the isomorphic packing,
\begin{equation}
\theta =  \theta_p ; \ \Omega_0 = - \sqrt{ \kappa^2 + \tau^2} .
\end{equation}
In this packing, the packing plane is fixed (say, the $x-y$ plane in the lab frame) and remains normal to the pitch axis of the helices ($\hat{z}$ direction).  Though filaments are parallel in the packing plane, inter-filament distances in this plane are not the true distances between the helical curves.  That is, $\uv$ and $\vv$ are not {\it relatively parallel} to the backbone, $\Tv$.  Below we show that this implies variation of inter-filament spacing the pitch axis of the helices in isomorphic bundles.

{\it Isometric Packing}: $\theta =0$ - In the isometric packing, filaments are parallel in the packing packing plane, $\pv = \Tv$ and there is no twist of the packing around $\Tv$.   Because $\uv$ and $\vv$ are relatively parallel to $\Tv$, distances in the packing plane are true inter-filament distances, and hexagonal packing in this plane implies uniform neighbor spacing along the length and throughout the cross section of the bundle.  While spacing is constant, shape varies throughout the packing plane according to
\begin{equation}
\delta \kappa = \kappa ; \  \delta \tau =\tau   ,
\end{equation}
which derive from eqs. (\ref{dkappa}) and (\ref{dtau}) for $\Omega_0 = -\tau$ and $\Omega_p = \kappa$, consistent with the shape variation of eq. (\ref{eq: kaptau}).

\subsection{Mechanics of filament distortion}

In this section, we describe the mechanical cost of distortions of filaments away from a preferred helical shape in bundles, characterized by constant values of backbone curvature $\kappa_0$ and torsion $\tau_0$.  We adopt the Kirchoff-Love theory of filament mechanics, where distortions are captured the rotation of a material frame along the backbone~\cite{landau}.  For each filament, labeled by position $\xv$, this frame is described by the triad $\{ \ev_1(\xv),  \ev_2(\xv),  \ev_3(\xv)\}$ where $\ev_3(\xv) = \Tv(\xv)$ is along the backbone tangent and 
\begin{eqnarray}
\nonumber
\ev_1(\xv) &=& \cos \phi (\xv)  \Nv(\xv) + \sin \phi (\xv) \Bv(\xv)  \\
\ev_2(\xv) &=&-  \sin \phi(\xv)  \Nv(\xv) + \cos \phi(\xv)  \Bv(\xv) ,
\end{eqnarray}
describe the position of the filament cross section, and $\phi$ represents an internal, rotational degree of freedom of the filament.  The geometry of filament backbone is locally described by three rotation rates,
\begin{equation}
\kappa_1 = \ev_1 \cdot \frac{ d \ev_3}{ d s} ; \ \kappa_2 = \ev_2 \cdot \frac{ d \ev_3}{ d s} ;  \ \omega = \ev_2 \cdot \frac{ d \ev_1}{ d s} ,
\end{equation}
where for simplicity we have suppressed the $\xv$ dependence of each quantity.  The first two rates, $\kappa_1 = \kappa \cos \phi$ and $\kappa_2 =  -\kappa \sin \phi$ describe the bending of the filament backbone projected onto the two material directions in the filament cross section, while the last rate describes the rotation, or torsion, of cross-section around the filament backbone, $\omega= \tau+ \partial_s \phi$.  In this framework, a helical filament is described by a preferred constant value of twist $\tau_0$, curvature $\kappa_0$, and a preferred direction of bending, which we choose to be $\ev_1$.  

For simplicity, we focus our attention on the limit of ``easy twist" filament mechanics~\cite{grason_pre_09}, which can be justified strictly for the case where the bending modulus of the filament $B$ is much larger than the twist modulus of the filament $C$.  In this limit, for any modulations of $\kappa(\xv)$ and $\tau(\xv)$ the torsional state of the filament adjusts to maintain alignment of the normal to $\ev_1$, the direction of preferred curvature.  Explicitly, $\phi= \partial_s \phi = 0$ and the mechanical cost of shape change for a filament at $\xv$ becomes simply,  
\begin{equation}
E_{\rm fil} (\xv) =\frac{1}{2} \int ds(\xv) \Big\{B\big[ \kappa (\xv) - \kappa_0\big]^2 + C \big[ \tau(\xv) - \tau_0\big]^2 \Big\}.
\end{equation}
To describe a filament bundle, we consider uniform areal density of filaments $\rho_0$ and integrate $E_{\rm fil} (\xv)$ over circular bundle cross section of radius $R$.  Using eq. (\ref{kaptau}) we find the total mechanical energy for a bundle $N$ filaments
\begin{multline}
\frac{E_{\rm mech}}{N L} \simeq \frac{B}{2} (\kappa- \kappa_0)^2  + \frac{C}{2} (\tau- \tau_0)^2  \\ + \frac{B}{8}  \delta \kappa^2 (\kappa R)^2+ \frac{C}{8}  \delta \tau^2 (\kappa R)^2 ,
\label{Emech}
\end{multline}
where $L$ is the filament contour length.  In deriving eq. (\ref{Emech}), we have assumed that shape of the central filament remains close the preferred shape for sufficiently narrow bundles, specifically, $|\kappa-\kappa_0| R \sim |\tau-\tau_0| R\sim ( \kappa_0 R)^3$, which follows from minimization of eq. (\ref{Emech}) with respect to $\kappa$ and $\tau$.  In the remainder of the article, we drop the distinction drop the distinct between the preferred helical shape and shape of the central filament, and take $\kappa = \kappa_0$ and $\tau = \tau_0$.  In Fig.~\ref{packing} we show graphically the spatial variation of bending energy density in bundles ranging from bundles ranging isometric to isomorphic packing.

\subsection{Thermodynamics of inter-filament packing}

We consider two models of the thermodynamic cost of inhomogeneous filament packing:  1) a continuum elastic model of cohesive filament bundles; and 2) the osmotic cost of excluded volume in densely packed bundles.  In both models, free energy costs arise as the consequence of the non-local nature of inter-filament contact.  Consider two filaments with backbones centered at $\xv$ and $\xv+\delta \xv$ in the packing plane.  Assuming that $\kappa a \ll 1$, filament backbones may be approximated locally as straight segments of orientation, $\Tv(\xv) \simeq \Tv(\xv+\delta \xv)$ for nearby points.  Rather than the in-plane separation $\delta \xv$, interactions between homogeneous filaments are natural functions of
\begin{equation}
{\bf \Delta}_\perp = \delta {\bf \xv} - \Tv ( \delta {\bf \xv}  \cdot \Tv) ,
\label{Delperp}
\end{equation}
which is the {\it distance of closest approach} between the filament centers~\cite{grason_pre_12}.  Since $|{\bf \Delta}_\perp|^2= | \delta {\bf \xv} |^2-(  \delta {\bf \xv} \cdot \Tv)^2$ when filaments are locally tilted into the packing plane, inter-filament spacing is necessarily reduced relative to the packing plane distance.  From eq. (\ref{dTperp}), the in-plane components  are
\begin{eqnarray}
\nonumber
T_{u} (\xv) &=& \sin ( \Omega s)  \sin \theta\big[ 1 + \cos \theta   \Omega_p (\Nv \cdot \xv)  \big] \\
T_{v} (\xv) &=& \cos ( \Omega s) \sin \theta  \big[ 1+ \cos \theta   \Omega_p (\Nv \cdot \xv)  \big]  ,
\label{Tin}
\end{eqnarray}
which shows that for $\theta \neq 0$ in-plane tilt (in the packing frame) varies along bundle, as well as within the bundle cross-section (for $\Omega_p \neq 0$).  The interfilament strain generated by tilt in the packing plane is depicted in Fig.~\ref{packing} graphically for bundles ranging isometric to isomorphic.  Below we find that inhomogeneous patterns of tilt, lead directly to thermodynamic costs associated with distortions away from the isometric state.

\subsubsection{Elastic cost of inter-filament distortions}

In this section we derive a model for filaments condensed by the finite-range attractive interactions, where the energetic cost of small changes in local inter-filament spacing $|\Delta_\perp|$ are described by a non-linear continuum elastic theory of columnar materials~\cite{selinger}.  The role of non-linear coupling between strains and tilts has been previously discussed and analyzed extensively in the context of inhomogeneous packing in chiral (twisted) filament bundles~\cite{grason_prl_07, grason_pre_09, grason_prl_10, grason_pre_12}.  The elastic energy has the form,
\begin{equation}
E_{\rm elast} = \frac{1}{2} \int dV \big[ \lambda (u_{kk})^2 + 2 \mu ~ u_{ij} u_{ij} \big] ,
\label{Eelas}
\end{equation}
where $\lambda$ and $\mu$ are the Lam\'e constants corresponding to the isotropic elastic response of a hexagonally-ordered material~\cite{selinger}.  Here, the 2-component strain, $u_{ij}$ describes changes in inter-filament spacing of nearby filaments, $|{\bf \Delta}_\perp|^2$, relative to initial separation $|{\bf \Delta}^0_\perp|^2$,
\begin{equation}
\delta \big( |{\bf \Delta}_\perp|^2\big) \simeq 2 u_{ij} \delta x_i \delta x_j ,
\end{equation}
where $\delta \xv$ is the two-component in-plane separation between filaments in the reference state.  We consider as a reference state, filaments initially oriented normally to the $(u,v)$ packing plane, so that indices $i$ and $j$ sum over $\uv$ and $\vv$ directions and $|{\bf \Delta}^0_\perp|^2= |\delta {\bf x}|^2$.   The form of the elastic strain follows,
\begin{equation}
u_{ij} \simeq \frac{1}{2} \big( \partial_i u_j + \partial_j u_i - T_i T_j \big) ,
\label{uij}
\end{equation}
which is sensitive only to changes in inter-filament spacing perpendicular to backbone orientation $\Tv$~\footnote{We have dropped a non-linear contribution to $u_{ij}$ proportional $\partial_i u_k \partial_j u_k$ because in-plane rotations are small~\cite{grason_pre_12}}.  Importantly, eq. (\ref{uij}) shows that inter-filament spacing, and consequently, inter-filament cohesive energy changes both with shifts of relative in-plane filament positions, as described by 
\begin{equation}
u^0_{ij} \equiv \frac{1}{2} (\partial_i u_j + \partial_j u_i),
\end{equation}
and with deformations that tilt backbones into the packing plane, as described by $ - T_i T_j/2$.   Note that while these latter terms vary along the bundles according to eq. (\ref{Tin}), {\it in the packing frame}  the relative positions of filaments are {\it fixed} along $s$, according to eq. (\ref{Rv}).   This implies that the net effect of in-plane tilt, when averaged over the length of the bundle, is to generate an {\it isotropic} compression, which in turn, generates a tensile response in the packing.  To offset the costs of tilt-induced compression, we consider an isotropic, in-plane strain by $u_{ij}^0 = ( \alpha + \beta r^2) \delta_{ij}$ 
where  $\alpha$ and $\beta$ are coefficients parameterizing the isotropic dilation of the packing at a radius $ r = |\xv|$ from the central filament.  Inserting into eq. (\ref{Eelas}) and minimizing over $\alpha$ and $\beta$ we find,
\begin{equation}
\frac{E_{\rm elast}}{V} \simeq  \frac{\mu}{8} \sin^4 \theta + O\big[ (\kappa R)^2 \big]
\label{Eelastic}
\end{equation}
where again we retain terms to lowest order in $\kappa R$ and $\rho_0 V /L =N$ where $\rho_0$ is the in-plane density of filaments.



\subsubsection{Osmotic cost of excluded volume }

In this section, we derive the cost associated with volume change of filaments bundled under the influence of an osmotic pressure, $\Pi$.  This model is relevant to the case of filaments in a solution of high-molecular weight polymers, whose size excludes them from the interstitial space between densely-packed filaments in condensed bundles.  The osmotic free-energy cost of a bundle is simply,
\begin{equation}
F_{\rm osm} = \Pi ~V_{\rm ex} ,
\end{equation}
where $V_{\rm ex}$ is volume within of the bundle.  Here, we consider a ``hard tube" model of filament packing, which requires that the local distance of closest approach between any filament and its neighbor in the packing satisfies,
\begin{equation}
|\bf{\Delta}_\perp| \geq d ,
\label{hard}
\end{equation}
where $d$ is the filament diameter.  As described above, local tilt of filaments into the packing plane frustrates the uniform spacing, hexagonal close-packing, requiring a distortion of the bundle cross-section.  To avoid overlap along nearest neighbor direction $\hat{\av}$ (unit vector in packing plane) we assume that the lattice spacing, $a$, in the packing plane {\it expands} to maintain,
\begin{equation}
d^2 = a^2 \big(1-[\hat{\av} \cdot \Tv(\xv) ]^2 \big) .
\end{equation}
Since the bond direction $\hat{\av}$ is  fixed in the packing-plane while in-plane components of $\Tv(\xv)$ rotate along the bundle according to eq. (\ref{Tin}), this condition must hold at $\xv$ for {\it all neighbor directions} $\hat{\av}$, including the in-plane tilt direction $\vv'$.  Neglecting the $O(\kappa r)$ correction to $\Tv(\xv)$ this gives us the packing-plane lattice dimension,
\begin{equation}
a = d \sec \theta ,
\end{equation}
with the areal density of filaments in the packing plane of $\rho(\theta) = \rho_0 \cos^2 \theta$, where $\rho_0 = 4/\sqrt{3} d^{-2}$.  The cross-sectional area of the packing plane is simply,
\begin{equation}
A_{\rm pack} = N \rho_0^{-1} \sec^2 \theta .
\end{equation}
The total volume excluded by the core of bundle is equal to the volume swept out by translating the cross-sectional area of the packing plane along the contour of the central filament ${\bf R}_0$.  Since the packing plane is inclined by an angle $\theta$ with respect to the tangent of ${\bf R}_0$, this excluded volume is simply,
\begin{equation}
V = A_{\rm pack} L \cos \theta ,
\end{equation}
and the osmotic cost forming the bundle becomes
\begin{equation}
\frac{F_{\rm osm}}{NL} = \Pi \rho_0 \sec \theta .
\label{Fosm}
\end{equation}
Notably, the volume-expansion cost for osmotically condensed filaments grows more quickly at small $\theta$ than for elastic distortion model.  

\section{Optimal packing and size of helical filament bundles}

We now analyze the structural transition of helical filament bundles of fixed radius in terms of the two models that penalize distortions of inter-filament packing, described by $E_{\rm elast}$ and $F_{\rm osm}$ derived in the previous section.  We determine the equilibrium variation of packing angle $\theta$, which serves as an order parameter, with bundle size, filament shape and respective costs of inter- and intra-filament distortion.  Specifically, we show that in the limit of strong packing forces, bundles approach the isometric state $\theta \to 0$ as radius vanishes, while in the opposite limit of large mechanical costs for intra-filament distortion, bundles approach the isomorphic state, $\theta \to \theta_p$.  

First, we analyze the state of packing in both models for a given $R$, $\kappa$ and $\tau$, and then consider the equilibrium size of freely-associating filament-bundles in solution.

\subsection{Cohesive bundles: elastically-constrained packings}

The free energy cost of filament distortion and inter-filament forces in cohesive bundles is described by
\begin{equation}
F_{\rm coh} = E_{\rm mech} + E_{\rm elast},
\end{equation}
where the forms of these functions are given be eqs. (\ref{Emech}) and (\ref{Eelastic}), respectively. Defining the dimensionless measures of curvature and torsion variation,
\begin{equation}
\delta \bar{\kappa}  \equiv \kappa^{-1} \delta \kappa ; \ \delta \bar{\tau}  \equiv \kappa^{-1} \delta \tau ,
\end{equation}
and grouping the $\theta$-dependent terms into $f_{\rm coh}(\theta)$ (free energy per filament)
\begin{multline}
\frac{f_{\rm coh} (\theta)} {  L} = \frac{B \kappa^4 R^2}{8} \Big\{| \delta \bar{\kappa} (\theta) |^2 +  \beta \cot^2 \theta_p  | \delta  \bar{\tau} (\theta)|^2 + \bar{R}_{\rm el}^{-2} \sin^2 \theta \Big\} ,
\label{fcoh}
\end{multline}
we note the state of packing (as described by $\theta$) is determined by following three dimensionless parameters for cohesive bundles
\begin{equation}
\tan \theta_p = -\kappa/\tau ; \ \beta = C/B; \ \bar{R}_{\rm el}= R/R_{\rm el} \ \Big( R_{\rm el} \equiv   \frac{ \sqrt{ \mu /(B \rho_0)} }{ 2 \kappa^2 } \Big)  .
\label{parcoh}
 \end{equation}
The first of these, $\tan \theta_p$, is a simply measure of central filament shape, while the second, $\beta$, is simply an intrinsic property of the mechanics of the filaments.  The last of these parameters, $\bar{R}_{\rm el}$, is a dimensionless measure of bundle size that characterizes the relative costs inter- vs. intra-filament distortions.  Hence, this parameter plays a role similar to the F\"oppl-von K\'arm\'an number for problems involving confined elastic membranes.  For small bundles, $\bar{R}_{\rm el} \ll 1$, the inter-filament elastic energies are large in comparison to the cost of filament shape change, while the inverse is true for large bundles, where $\bar{R}_{\rm el}  \gg 1$.  

\begin{figure}
\epsfig{file=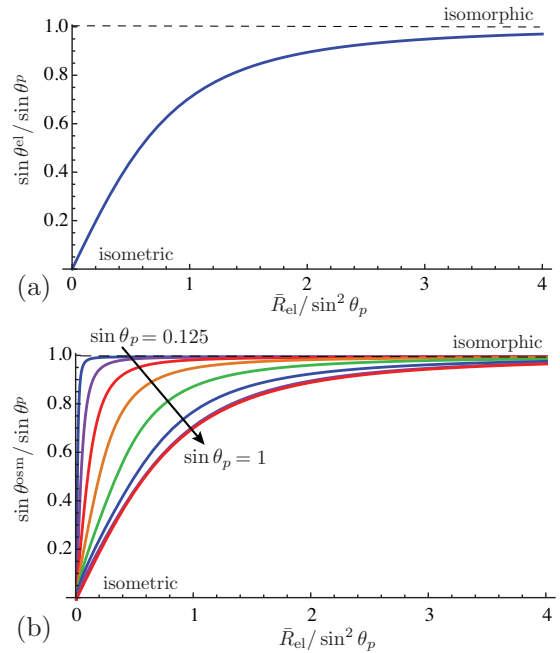, width=2.85in}
\caption{In (a), the variation of optimal packing angle for cohesive bundles with radius of the ``easy twist" limit ($\beta \to 0$).  The finite torsional stiffness case ($\beta =1$) is shown in (b), for a sequence of packing angles: $ \sin \theta_p = 0.125, 0.25, 0.375, 0.5,0.625, 0.75, 0.875, 1$.  }
\label{the_el}
\end{figure}

To understand the size-dependence of packing we consider two cases for $\beta$.  First, we analyze the extreme ``easy twist" limit, where $\beta \to 0$, and the minimization of $f_{\rm coh} $ with respect to $ \sin \theta$ is analytically tractable, resulting in
\begin{eqnarray}
\nonumber
 \sin\theta_{\rm coh} (\beta \to 0) &=& \Big( \frac{ \sin^2 \theta_p \bar{R}_{\rm el} ^2 }{\sin^4 \theta_p + \bar{R}_{\rm el} ^2 } \Big)^{1/2}\\ &\simeq & \left\{ \begin{array}{ll} \bar{R}_{\rm el} / \sin \theta_p & {\rm for} \ \bar{R}_{\rm el}  \ll \sin^2 \theta_p \\
\sin \theta_p & {\rm for} \ \bar{R}_{\rm el}  \gg \sin^2 \theta_p \end{array} \right.
\label{thecob0}
\end{eqnarray}
As shown in Fig.~\ref{the_el} (a) this behavior demonstrates that cohesive bundles adopt the extremal isometric or isomorphic packings only in the $R \to 0$  (single filament) and $R \to \infty$ (bulk assembly) limits where the respective costs of inter- and intra-filament distortion dominate.

Extending the analysis to the case of a finite twist elastic cost, $\beta \neq 0$, we consider the small $\sin \theta$ limit of eq. (\ref{fcoh}), finding optimal packing for narrow bundles,
\begin{equation}
\lim_{ \bar{R}_{\rm el} \to 0}  \sin \theta_{\rm coh} (\beta) \simeq \big( \bar{R}^2_{\rm el}  \beta \cot^3 \theta_p \big/2)^{1/3},
\end{equation}
which indicates a much more rapid change in packing angle with size for small bundles, $\sim R^{2/3}$, due to finite torsional stiffness.  The change in power-law dependence derives from the fact that in the small angle limit the $\delta \kappa^2 (\theta) - \delta \kappa^2 (0) \sim - \sin^2\theta$, whereas  $\delta \tau^2 (\theta) - \delta \tau^2 (0) \sim - \sin\theta$.  Hence, the ``torque" acting on packing angle generated by bending vanishes in the isometric state ($\theta =0$), while the torque generated by torsional cost remains finite for finite $\beta \cot^3 \theta_p$.    The singular, power-law evolution of  $\sin \theta_{\rm coh} (\beta)$ at small bundle size is also strongly dependent on filament geometry.  As filament geometry approaches shape, $\theta_p \to 0$, the small-$R$ evolution from isometric to isometric bundles becomes infinitely sharp, indicative of the negligible disruption of inter-filament packing in this limit.  In the opposite limit of large tilt angles $\theta_p \to \pi/2$ the cost of filament torsion in the isometric state becomes negligible, and the dependence of packing angle on bundle size (for narrow bundles) approaches the linear evolution of eq. (\ref{thecob0}).  The numerical solution for $\theta_{\rm coh} (\beta=1)$ is shown in Fig.~\ref{the_el} (b) for several filament shapes, ranging from the rapid jump between isometric to isomorphic bundles for nearly straight filaments  ($\kappa/ \tau \ll1$) to the more gradual evolution of packing in highly curved filaments ($\kappa/ \tau \gg1$).

\subsection{Osmotic bundles: pressure-constrained packings}

The fre- energy cost of distortion of filament shape and inter-filament forces in osmotically-condensed bundles is described by
\begin{equation}
F_{\rm osm} = E_{\rm mech} + F_{\rm osm},
\end{equation}
where the forms of these functions are given be eqs. (\ref{Emech}) and (\ref{Fosm}), respectively.  Following a similar analysis to the above for cohesive bundles we group the $\theta$-dependent terms into $f_{\rm osm}(\theta)$ 
\begin{multline}
\frac{f_{\rm osm} (\theta)} {   L} = \frac{B \kappa^4 R^2}{8} \Big\{| \delta \bar{\kappa} (\theta) |^2 +  \beta \cot^2 \theta_p | \delta  \bar{\tau} (\theta)|^2 + 4\bar{R}_{\rm os}^{-2} \sec \theta \Big\} .
\label{fosm}
\end{multline}
For pressure-induced bundles, the packing is determined by following three dimensionless parameters
\begin{multline}
\tan \theta_p = -\kappa/\tau ; \ \beta = C/B; \\ \bar{R}_{\rm os} = R/R_{\rm os} \ \Big( R_{\rm os} \equiv \sqrt{2  \Pi/( B \kappa^4 \rho_0)}\Big)\  .
\label{parosm}
 \end{multline}
The first of these two parameters, reflecting intrinsic filament properties, are appear identically to the analysis of cohesive bundles eq. (\ref{parcoh}), where as the dimensionless measure of bundle size, $\bar{R}_{\rm os}$, captures the relative cost of bundle volume change to filament distortion, including the dependence on osmotic pressure.

\begin{figure}
\epsfig{file=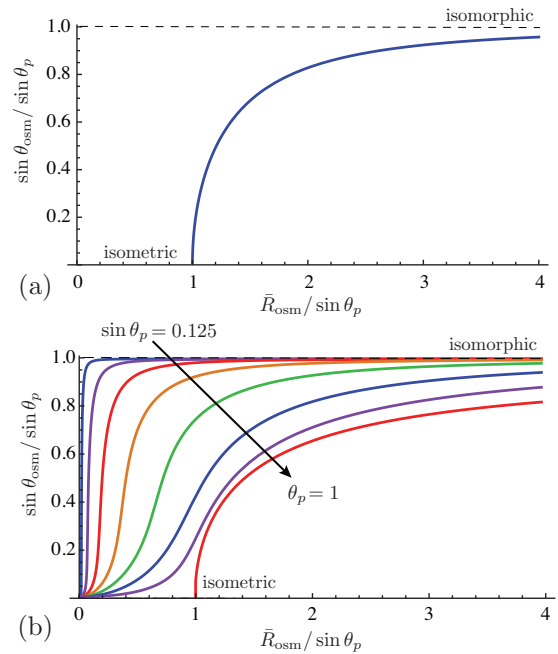, width=2.85in}
\caption{In (a), the variation of optimal packing angle for osmotically-condensed bundles with radius of the ``easy twist" limit ($\beta \to 0$), shown here for $\sin \theta_p = 0.5$.  The finite torsional stiffness case ($\beta =1$) is shown in (b), for a sequence of packing angles: $ \sin \theta_p = 0.125, 0.25, 0.375, 0.5,0.625, 0.75, 0.875, 1$.}
\label{the_osm}
\end{figure}

Again, we first analyze equilibrium packing in the $\beta \to 0$ limit.  The minimization of $f_{\rm osm}$ is further simplified by noting $\sin^2 \theta < 1$ which allows us to approximate $\sec \theta \simeq 1+ \sin^2 \theta /2 + 3 \sin^4 \theta /8$.  Minimizing with respect to $\theta$ in this limit we find,
\begin{equation}
\sin^2 \theta_{\rm osm} (\beta \to 0) = \left\{ \begin{array}{ll} 0  & {\rm for} \ \bar{R}_{\rm os} \leq \sin \theta_p \\
\frac{ \sin^2 \theta_p (\bar{R}_{\rm os} ^2 - \sin^2 \theta_p) }{\bar{R}_{\rm os} ^2 + 3\sin^4 \theta_p /2 } & {\rm for} \ \bar{R}_{\rm os} > \sin \theta_p \end{array} \right.
 \end{equation}
which is shown in Fig. \ref{the_osm} (a).  Thus, in comparison to cohesive bundles, we find a more abrupt change a packing with bundle size for osmotic bundles.  Notably we find a {\it second-order transition} from the isometric state ($\theta =0$) occurring at a critical size $\bar{R}_{\rm os} = \sin \theta_p$.  Near to this critical value, for $\bar{R}_{\rm os} \gtrsim \sin \theta_p$, we find $\theta_{\rm osm} \sim |\bar{R}_{\rm os} -\sin \theta_p|^{1/2}$ indicating a high sensitivity of packing near the critical size.  As in the case of elastic restoring forces, we find that large bundles $\bar{R}_{\rm os} \gg  \sin \theta$ asymptotically approach the isomorphic packing $\theta_{\rm osm} \to \theta_p$.

Following the analysis of finite torsional stiffness ($\beta \neq 0$) above for cohesive bundles, we analyze the small $\sin \theta$ limit of eq. (\ref{fosm}) to determine the small diameter evolution of packing,
\begin{equation}
\lim_{ \bar{R}_{\rm os} \to 0}  \sin \theta_{\rm osm} (\beta) \simeq \bar{R}^2_{\rm os}  \beta \cot^3 \theta_p/2 .
\end{equation}
Hence, for finite torsional stiffness, bundles are distorted from the isometric packing for any non-zero bundle radius, deriving again from the non-zero torque exerted on the isometric bundle by torsional mechanics of filaments.  Fig. \ref{the_osm} (b) shows the evolution of packing angle with bundle radius for $\beta =1$.  Notably, as we found for cohesive bundles, when the helical angle approaches $\theta_p \to \pi/2$, effects of torsional stiffness at small packing angles vanish, and we find that bundles remain isometric for a finite range of small bundles, exhibiting a second-order transition to $\theta_{\rm os} \neq 0$ at a critical size $\bar{R}_{\rm os}$.

\subsection{Self-limited bundle size}

Having analyzed the size-variation of optimal packing, we now consider the equilibrium size of bundles formed in concentrated filament solutions.  In a previous study of helical filament bundles~\cite{grason_pre_09}, which assumed the packing remains locked to the isometric limit, it was shown that the thermodynamically optimal size of bundles remains finite over a broad range of cohesive energy due to the prohibitive cost of filament shape change in large bundles.  Presently, we generalize this analysis by considering the relaxation of filament distortion through adjustment of bundle from the constant-spacing towards constant-shape packing.  

We consider solutions in the canonical ensemble, where the filament number and volume are constant.  We further assume that all but a negligible number of filaments belong to bundles (i.e. far above the saturation point) and consider a size distribution that is sharply peaked around radius $R$.  In this case, an additional-free energy cost enters, associated with the surface energy, $\Sigma$, and the exposed bundle surface area ,
\begin{equation}
F_{\rm surf} = \Sigma 2 \pi R L ,
\end{equation}
where we assume $L\gg R$ such that the contribution from the bundle ends is negligible. 

To analyze the evolution of bundle size distribution with $\Sigma$, we assume $|\delta \kappa R|$ and $|\delta \tau R|$ remains sufficiently small that the central filament locks into the preferred filament geometry, $\kappa = \kappa_0$ and $\tau_0 = \tau$. For the purposes of simplifying the analysis we focus on to the limit of easy twist where $\beta =0$, though it is straightforward to extend the more general analysis of $\beta \simeq 1$.   The free energy per filament of bundles of size $R$ may be written as,
\begin{equation}
\frac{ F_{\rm tot}(R)}{ N L } = \frac{ 2\Sigma \rho_0^{-1}}{R} +\frac{ F_{\rm bulk} (R)}{ N L} ,
\end{equation}
where $F_{\rm bulk}$ represents the bulk costs associated with packing distortion (optimized over packing angle, $\theta$).  

\begin{figure}
\epsfig{file=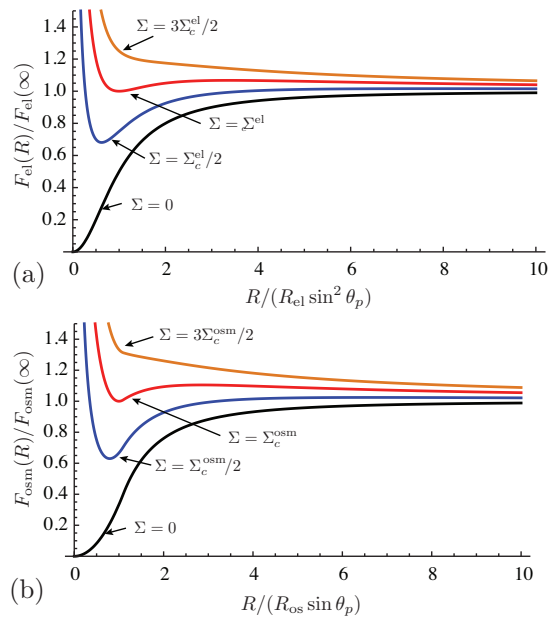, width=2.85in}
\caption{In (a) and (b), the respective free energy of cohesive and osmotically-condensed bundles of helical filaments ($\sin \theta_p =1/2$) as function of bundle radius for a range of surface energies as indicated in the plots.  Both condensation mechanisms show finite-radius equilibrium below a threshold surface energy.}
\label{FvsR}
\end{figure}



In both models, the cost associated with the competition between inter- and intra-filament packing are monotonically increasing functions of $R$ (see Fig.~\ref{FvsR}).  For small radius --- $R \ll R_{\rm el} \sin^2 \theta_p$ for cohesive bundles and $R \leq R_{\rm os} \sin \theta_p$ for osmotic bundles --- bundles are packed isometrically and the mechanical cost associated with filament distortion grows as $F_{\rm bulk}/(N L) \sim  \kappa^4 R^2$.  In the opposite large bundle limit, bulk cost of packing in both bundles (per filament) saturates as the bundle approaches the isomorphic state $\theta \to \theta_p$, where $F^{\rm elas}_{\rm bulk}/(N L) = \mu\rho_0^{-1}  \sin^4 \theta_p$ and $F^{\rm elas}_{\rm bulk}/(N L) =  \Pi \rho_0^{-1} \sec \theta_p$.  Hence, $F_{\rm bulk}/(N L)$ favors small bundles, acting against lateral bundle growth.    

As shown in Fig.~\ref{FvsR} , below a critical value of surface energy, $\Sigma_c$, the minimal free energy state is characterized by a finite bundle size.  The critical surface energy derives from the solution to 
\begin{equation}
f_{\rm bulk} (\infty) -  f_{\rm bulk} (R_c) = R_c \frac{ d f_{\rm bulk}}{ d R}\Big|_{R_c} = L \frac{ 2 \Sigma_c \rho_0^{-1} }{R_c} ,
\end{equation}
where $f_{\rm bulk} (R_c) = F_{\rm bulk}/N$ and $R_c$ is finite-size of bundles at the point of equilibrium with the bulk assembly ($R \to \infty$).  For cohesive bundles this yields
\begin{equation}
R^{\rm el}_c= R_{\rm el} \sin^2 \theta_p ; \ \Sigma_c^{\rm elas}= \Sigma_c^{\rm elas}  \sin^6 \theta_p   \ \bigg(\Sigma_c^{\rm elas} \equiv \frac{B \kappa^4  R_{\rm el}^3 \rho_0 }{ 32}  \bigg)  ,
\label{sigc_el}
\end{equation}
while for osmotic bundles, it gives
\begin{eqnarray}
R^{\rm os}_c&=& 2 R_{\rm os} \sqrt{(\sec \theta_p -1)/3 }  \ {\rm for } \ R^{\rm os}_c < R_{\rm os}   \sin \theta_p  \\ \nonumber
\Sigma_c^{\rm osm}&=& (\sec \theta_p -1 )^{3/2} \ \bigg(\Sigma_c^{\rm osm} \equiv \frac{B \kappa^4  R_{\rm os}^3 \rho_0 }{  3^{3/2} } \bigg)  .
\label{sigc_os}
\end{eqnarray}
When $R^{\rm os}_c < R_{\rm os} \sin \theta_p$ the optimal bundle remains isometric at the critical surface energy, which holds only for $\theta_p \leq 0.71$.  For larger helix angles, the packing and size of osmotic bundles must be determined numerically.  We may estimate the thermodynamic properties in this limit, however, by approximating equilibrium bundle size to be at the critical value, $R_{\rm iso}= R_{\rm os} \sin \theta_p$.  Since $F_{\rm tot} (R_{\rm os} \sin \theta)$ this overestimates the free energy of the finite-size bundle, this approximation gives a lower bound on the critical surface energy,
\begin{equation}
\Sigma_c^{\rm osm} > \frac{\rho_0}{2} R_{\rm os} \sin \theta_p\big[ f_{\rm bulk} (\infty) - f^{\rm bulk}  \big( R_{\rm os} \sin \theta_p \big) \big] .
\label{sigc_os_large}
\end{equation}

\begin{figure}
\epsfig{file=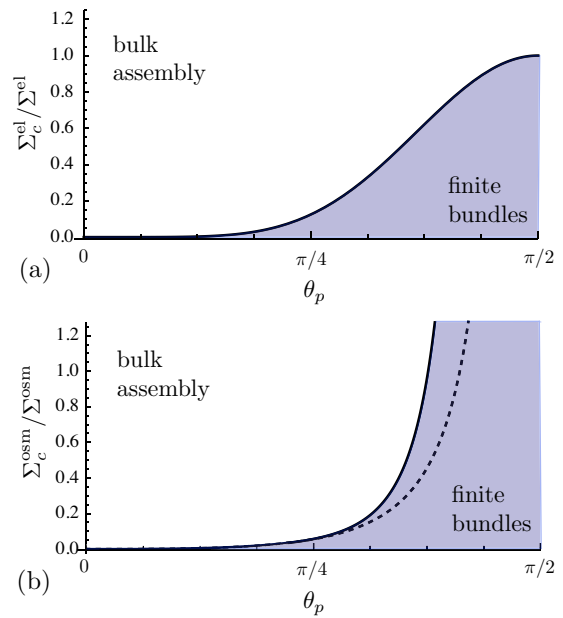, width=2.85in}
\caption{The phase diagrams of cohesively-driven and osmotically-driven assembly of helical filaments, respectively in (a) and (b), in terms of helical angle of filaments and surface energy bundles.  Regions of self-limited (bundles) and bulk assembly are indicated.} 
\label{phase}
\end{figure}

In Figure \ref{phase} we plot the phase diagrams for models of cohesion- and pressure-driven condensation of helical filaments, where a critical surface energy $\Sigma_c$ separates a phase of finite-size bundles from bulk assembly (in the isomorphic packing).  From eqs. (\ref{sigc_el}) and  (\ref{sigc_os}) it is easy to show that for small helix angles that critical surface energy increases as power law for both models:  $\Sigma^{\rm elas}_c \sim \theta_p^6$ and $\Sigma^{\rm osm}_c \sim \theta_p^3$.  This gives the intuitive result that as the preferred filament approaches a straight geometry, thermodynamic costs associated with packing frustration must vanish.  Minimization of the surface energy (for any value of $\Sigma$) for bundles of rod-like filaments results in aggregates of unlimited lateral size, the result of standard arguments for 2D aggregation.  In the large helix angle limit, the costs of inter-filament packing of the isomorphic, $R \to \infty$ limit  increase in both models.  Consequently, the critical surface energy needed to offset the bulk costs of isomorphic state also increases indicating a broader thermodynamic stability of finite-sized bundles.  Hence for cohesive bundles we find $\Sigma_c^{\rm el}\sim R^{\rm elas}_c \mu \sin^4 \theta_p \sim \sin^6 \theta_p$, and a maximal stability for bundles at $\theta_p \to \pi/2$, or the limit $\kappa\gg \tau$.  We find a profoundly different result for the $\theta_p \to \pi/2$ limit of osmotically-driven bundles, as $\Sigma^{\rm osm}_c$ diverges due to divergent cost of isomorphic packing of the ``ring-filament" limit.

\section{Discussion \& Conclusion}

In this paper, we studied a continuum model for the formation of bundles of intrinsically-curved, helical filaments.  A key aspect of this problem is a fundamental frustration between the curved shape of individual filaments and the tendency for even filament spacing, or maximal density, in the cross section.  As implied by eq. (\ref{eq: kaptau}), isometric packing requires filament distortions, whose costs increase rapidly with cross-section thickness.  In comparison, the thermodynamic costs of imperfect packing in both models is relatively insensitive to bundle radius.  While the (per filament) cost of filament distortion in isometric bundles grows as $\sim \kappa^4 R^2$, according to eqs. (\ref{Eelastic} ) and (\ref{Fosm}) the inter-filament packing cost in isomorphic (per filament) is independent of radius.  This difference in size sensitivity underlies the generic evolution from isometric packings for small bundles ($R \to 0$), where mechanical distortion is negligible, to isomorphic packings for large bundles ($R \to \infty$), where filament distortion becomes prohibitively expensive.  

Despite the common tendency towards constant shape packings in larger bungles, we find a markedly different sensitivity of inter-filament packing to lateral bundle radius for cohesive bundles and osmotically-condensed bundles, as evidenced by the $R$-dependence of optimal packing angles shown in Figs. \ref{the_el} and \ref{the_osm}.  In cohesive bundles, packing evolves smoothly with $R$ for all helix angles and mechanical properties, exhibiting a non-isometric packing (i.e. $\theta_{\rm el} \neq 0$) for any finite radius.  In contrast, when the cost of filament torsion is negligible (either $\beta \to 0$ or $\theta_p \to \pi/2$), in osmotically-condensed bundles, we find that bundles ``lock-in" to the isometric packing over finite range of small radii, $R < R_{\rm osm} \sin \theta_p$, exhibiting a continuous, second-order transition to the non-isometric packing beyond this critical size.  The difference in $R$-dependence of the packing in the two models derive ultimately from a different ``stiffness" to interfilament distortion from the isomorphic state, $\delta(|{\bf \Delta}_\perp|) \sim -\theta^2$.  In the elastic model this leads directly to a packing angle dependence, $F_{\rm elas}\sim \big| \delta(|{\bf \Delta}_\perp|) \big|^2\sim \theta^4$, indicating a ``soft" $\theta$-dependence of packing cost that is generically weaker than the $\theta$-dependence of $E_{\rm mech}$ driving bundles toward the $\theta \to \theta_p$ isomorphic state.  This can be contrasted with the ``stiffer" $\theta$-dependance of packing free energy in osmotic bundles, $F_{\rm elas}\propto \Delta V \sim \big| \delta(|{\bf \Delta}_\perp|) \big| \sim \theta^2$, a comparable dependence to the bending costs, which implies a finite size range over which the isometric packing is stable.  

The evolution from isometric to isomorphic packing has important consequences for the equilibrium size of self-assembled bundles of helical filaments.  In ref. \cite{grason_pre_09}, it was shown that constraining bundles of helical filament to the isometric packing leads to self-limited lateral radii of equilibrium bundles.  Hence, this system belongs to an unusual class of associating molecular systems along with self-twisted bundles of chiral filaments~\cite{grason_prl_07, grason_pre_09, turner, hagan_10, heussinger}, where the equilibrium dimensions of the condensed state are finite, but mesoscopic --- much larger that filament diameter --- despite the finite range of the interactions. Importantly, the present study demonstrates that even when inter-filament packing is free to relax from the isometric packing, the self-limiting behavior of helical filament bundles is retained over a broad range of cohesive conditions.  From Fig. \ref{phase}, we note again a significant difference between the stable range of self-limited bundles in cohesive and osmotic bundles, characterized the large-$\theta_p$ limit of the $\Sigma_c$, the surface energy where finite-size bundles are in equilibrium with bulk (infinite-$R$) assembly.  While finite-diameter cohesive bundles reach a maximal range of stability in ring-filament limit, $\Sigma^{\rm el}_c(\theta_p \to \pi ) = \Sigma^{\rm el}$, $\Sigma^{\rm osm}_c$ diverges in this limit, indicating that the stable range of finite bundles becomes infinite in the $\kappa/ \tau \gg1$ limit, where the costs of packing frustration are maximal.  

We conclude with a brief discussion of relevant physical parameters for a model system of helical filaments, the bacterial flagella.  Owing the precise control of helical geometry offered through the multiple polymorphic shapes of flagella~\cite{asakura_oosawa_book, darnton_berg}, the condensed states of reconstituted flagella provide an ideal system for exploring the thermodynamics of curved-filament assembly.  Recent experiments have exploited the intrinsic shape of reconstituted flagella in colloidal suspensions as a means to study study the unusual liquid crystalline mesophases of spiral-shaped filaments~\cite{berry}.  Here, we consider the properties of {\it bundles} of flagella that may be formed in highly-concentrated solutions based on our analysis.   For simplicity, we focus on the case of bundles of flagella osmotically-condensed in the presence of a osmolytes sufficiently large to be excluded from the interstitial regions of the dense bundle core.  Of particular importance is the size scale, $R_{\rm os}$ in eq. (\ref{parosm}), which determines both the range over which bundles are isometrically packed ($R \leq R_{\rm os} \sin \theta_p$ for $\beta =0$) as well as size range of finite-size bundles in equilibrium according to eq. (\ref{sigc_os}).  We recast the definition of this length as $R_{\rm os} = d (\Pi/ \Pi_{\rm mech} )^{1/2}$, where 
\begin{equation}
\Pi_{\rm mech} \approx B \kappa^4 ,
\end{equation}
constitutes a pressure scale determined by the deformation cost of filaments (we have dropped numerical factors of order unity).  Force-extension measurements of flagella under flow suggest a bending modulus, $B = 3.5 ~{\rm pN/ \mu m^2}$~\cite{darnton_berg}.  Using the preferred curvature of the {\it normal} and {\it coiled} polymorphs, $\kappa_{normal}= 1.3 ~{\rm \mu m}^{-1}$ and $\kappa_{coiled}= 1.8 ~{\rm \mu m}^{-1}$, we find predictions for the characteristic osmotic pressures, $\Pi(normal) \approx 10~{\rm Pa}$ and $\Pi(coiled) \approx 38~{\rm Pa}$.  Compared to the typical scale of osmotic pressure --- of order ~${\rm atm} \approx 10^5~{\rm Pa}$ for few percent of 8000 MW PEG --- the modest pressure scale set by flagellar bending cost suggests that the characteristic bundle size is {\it mesoscopic}, orders of magnitude larger filament diameter, $R_{\rm osm} \sim10^2 d$.  Since $\sin \theta_p \approx 1$ for helical angles of these flagellar polymorphs --- $\theta_p(normal) \simeq 31^\circ$ and $\theta_p(coiled) \simeq 77^\circ$ --- our theory suggests that packings remain isomorphic for {\it bundle radii up to the range of microns}.  

Turning now to the range of self-limited lateral assembly, we consider the surface energy of bundle under the influence of osmotic pressure, $\Sigma = \Pi r$, where $r$ is the effective size of that ``depletion zone" (roughly, the osmolyte radius) that excludes osmolytes near the bundle surface~\cite{marenduzzo}.  From the critical surface energy of eq. (\ref{sigc_os}), we expect finite diameter bundles {\it above} a critical pressure, $\Pi_c \approx (r/d)^2 \Pi_{\rm mech} (\sec \theta_p -1)^{-3}$, again suggesting the threshold osmotic pressures are nominal, in the range the $1-10$ Pa.  Thus, we expect reasonable experimental conditions to be in the range of self-limited bundle sizes, for which the balance between surface energy and filament distortion in isometric bundles determines the equilibrium dependence of bundle size on osmotic pressure,
\begin{equation}
R /d \approx (\Pi/\Pi_{\rm mech})^{1/3}  ,
\end{equation}
which is of order 10 for $\Pi \approx 1~{\rm atm}$.  These estimates suggest a strong (measurable) pressure dependence self-limited diameter of flagellar bundles in the range of weak osmotic pressures $\Pi \gtrsim \Pi_{\rm mech}$.


\begin{acknowledgments}
I would like to thank L. Cajamarca for numerous discussions and a careful reading of the manuscript.  I would also like to acknowledge P. Ziherl and the hospitality of the Jo\u{z}ef Stefan Institute at the University of Ljubljana where some of this work was completed.  I am grateful to Z. Dogic and S. Yardimci for providing the florescence image of flagellar bundling.    This work was supported by the NSF through grant CMMI 10-68852.
\end{acknowledgments}

\begin{appendix}

\section{Derivation of Shape Variation} 
Here, we derive the shape variation of filaments in the family of bundles described by eqs. (\ref{Rv})-(\ref{omegas}).  The unit tangent, $\Tv(\xv)$, of filaments at $\xv$ follows straightforwardly from eq. (\ref{Rv})
\begin{equation}
\Tv(\xv) = \bigg| \frac{ \partial s(\xv) }{ \partial s} \bigg|^{-1} ( \Tv + \delta \Tv) \simeq \Tv (1-\Tv \cdot \delta \Tv) + \delta \Tv ,
\end{equation}
where 
\begin{equation}
\delta \Tv \equiv  \partial_s \xv = ( \Omega +\Omega_0) \pv \times \xv -  \Omega_p (\Nv \cdot \xv) \pv  ,
\label{dT}
\end{equation}
and
\begin{equation}
(\delta \Tv \cdot \Tv) =  \big[ (\Omega+ \Omega_0)\sin \theta - \cos \theta \Omega_p\big] (\Nv \cdot \xv)  .
\label{dT}
\end{equation}
\begin{equation}
\frac{ \partial s(\xv) }{ \partial s} = | \Tv + \delta \Tv| \simeq 1 + \Tv \cdot \delta \Tv ,
\end{equation}
reflects mapping of arc-length elements at $\xv$ to the central filament arc-length.  Recall in this notation objects that appear as explicit functions of $\xv$ denote geometry quantities at position $\xv$ in the packing plane, and objects not written as functions of $\xv$ refer to geometric quantities related to the central filament at $\xv =0$. Note also that we are working in the limit of narrow bundles where it is sufficient to consider first-order in $|\xv|$ correction to the geometry (relative to the central filament).

The filament curvature derives from the magnitude of 
\begin{equation}
\frac{d \Tv(\xv)}{d s(\xv)} \simeq  \frac{ \kappa \Nv (1-\Tv \cdot \delta \Tv ) - \Tv \partial_s (\Tv \cdot \delta \Tv ) + \partial_s \delta \Tv  }{1 + \Tv \cdot \delta \Tv}  ,
\end{equation}
from which we derive
\begin{equation}
\kappa(\xv) - \kappa \simeq\Nv \cdot \partial_s \delta \Tv - 2 \kappa (\Tv \cdot \delta \Tv) ,
\label{kappaA}
\end{equation}
where
\begin{multline}
\partial_s \delta \Tv \simeq - (\Omega+\Omega_0)^2 \xv +\Omega_p (\Omega+\Omega_0) \uv' \times \xv \\ + \Omega_p \Omega (\vv' \cdot \xv) \pv - \Omega_p^2 (\uv' \cdot \xv) \uv' .
\label{dsdT}
\end{multline}
Using the fact that $\uv' = \Nv$ is perpendicular to $\pv$ we find,
\begin{equation}
\Nv \cdot \partial_s \delta \Tv = -\big[(\Omega+\Omega_0)^2 + \Omega_p^2\big] (\Nv \cdot \xv) ,
\end{equation}
which when substituted into eq. (\ref{kappaA}) yields eq. (\ref{dkap}). 
\begin{multline}
\delta \kappa = -\kappa^{-1} \big[ (\Omega+ \Omega_0)^2 + \Omega_p^2 \big] 
-2 \big[ (\Omega+ \Omega_0)\sin \theta - \cos \theta \Omega_p\big]  ,
\label{dkappaA}
\end{multline}
where $\kappa(\xv)-\kappa = \delta \kappa (\kappa\Nv \cdot \xv)$.

The variation of torsion follows from
\begin{equation}
\tau(\xv) = - \Nv(\xv) \cdot \Big( \frac{d \Bv(\xv)}{d s(\xv)} \Big) \simeq \tau(1-\Tv \cdot \delta \Tv ) - \Nv \cdot \partial_s \delta \Bv_\perp   ,
\label{taueq}
\end{equation}
where $\delta \Nv_\perp \simeq \Nv(\xv)-\Nv$ and  $\delta \Bv_\perp \simeq \Bv(\xv) -\Bv$ are the first order variation to normal and bi-normal, respectively, relative to the central filament.  The normal correction follows from,
\begin{eqnarray}
\nonumber
\delta \Nv_\perp &=& \frac{1}{\kappa(\xv)} \frac{d \Tv(\xv)}{d s(\xv)} - \Nv \\ &\simeq& \kappa^{-1} (\Bv \cdot \partial_s \delta \Tv) \Bv - \Tv (\Nv \cdot \delta \Tv),
\end{eqnarray}
where we used $\Nv(\xv) \cdot \Tv(\xv) = O(|\xv|^2)$ in the second line.  Similarly, defining $\delta \Tv_\perp = \delta \Tv - \Tv(\Tv \cdot \delta \Tv),$  we have the correction to the bi-normal,
\begin{eqnarray}
\nonumber
\delta \Bv_\perp &\simeq& \Tv \times \delta \Nv_\perp + \delta \Tv_\perp \times \Nv \\
&=& - \Nv (\Bv \cdot \partial_s  \delta \Tv) \kappa^{-1} -\Tv ( \Bv \cdot \delta \Tv ) .
\label{dB}
\end{eqnarray}
The $\Nv$ component of the derivative of eq. (\ref{dB}) can be written as
\begin{equation}
\Nv \cdot \partial_s \delta \Bv_\perp =\frac{ \tau}{\kappa} (\Nv \cdot \partial_s \delta \Tv) - (\Bv \cdot \partial^2_s \delta \Tv)\kappa^{-1} - \kappa (\Bv \cdot \delta \Tv) .
\end{equation}
From eq. (\ref{dT}) we have
\begin{equation}
(\Bv \cdot \delta \Tv)=\big[ \cos \theta (\Omega+\Omega_0)+\sin \theta \Omega_p\big] (\Nv \cdot \xv), 
\end{equation}
and differentiating eq. (\ref{dsdT}) we find
\begin{multline}
(\Bv \cdot \partial_s^2 \delta \Tv) = \Big\{-(\Omega+\Omega_0)^2 \big[ \cos \theta (\Omega+\Omega_0)+\sin \theta \Omega_p\big]  \\  -\Omega_p \Omega (\Omega+\Omega_0)\sin \theta  -\sin \theta  \Omega_p\Omega^2  -\Omega_p^2\tau \Big\} (\Nv \cdot \xv) .
\end{multline}
Substituting these into eq.~(\ref{taueq}) we find
\begin{multline}
\delta \tau =- \frac{\tau}{\kappa} \big[ (\Omega+ \Omega_0)\sin \theta - \cos \theta \Omega_p\big]    -\kappa^{-2}\Big\{ - \tau \big[ (\Omega+ \Omega_0)^2 + \Omega_p^2 \big] \\ + \big[ \kappa^2+ (\Omega+ \Omega_0)^2 \big] \big[ \cos \theta (\Omega+\Omega_0)+\sin \theta \Omega_p\big]  \\ - \Omega_p\big[\sin \theta (\Omega+\Omega_0) \Omega + \sin \theta \Omega^2 + \Omega_p \tau \big] \Big\}  ,
\label{dtauA}
\end{multline}
where  $\tau(\xv)-\tau =  \delta \tau (\kappa\Nv \cdot \xv)$.

\end{appendix}

\end{document}